\documentclass[12pt]{article}
\usepackage{a4wide}
\usepackage{graphicx}
\usepackage{latexsym}

\usepackage{axodraw}

\newcommand{\SMN}{\overline \Pi_{Sij}^M}

\newcommand{\vl}{\overline \Pi_{Vij}^{(0)}}

\newcommand{\vlt}{\overline \Pi_{Vij}^{(0+1)}} 
\newcommand{\vltc}{\overline \Pi_V^{(0+1)\chi}}
\newcommand{\vlti}{\overline \Pi_V^{(0+1)I}}
\newcommand{\vt}{\overline \Pi_{Vij}^{(1)}}

\newcommand{\s}{\overline \Pi_{Sij}}

\newcommand{\vlf}{\Pi_{Vij}^{(0)}}
\newcommand{\vtf}{\Pi_{Vij}^{(1)}}
\newcommand{\smnf}{\Pi_{Sij}^M}      
\newcommand{\ssf}{\Pi_{Sij}}
\newcommand{\PMN}{\overline \Pi_{Pij}^M}
\newcommand{\al}{\overline \Pi_{Aij} ^{(0)}}
\newcommand{\alt}{\overline \Pi_{Aij}^{(0+1)}} 
\newcommand{\altc}{\overline \Pi_A^{(0+1)\chi}}
\newcommand{\alti}{\overline \Pi_A^{(0+1)I}}
\newcommand{\at}{\overline \Pi_{Aij}^{(1)}}

\newcommand{\PMNc}{\overline \Pi_{P}^{M\,\chi}}
\newcommand{\PMNi}{\overline \Pi_{P}^{M\,I}}
\newcommand{\p}{\overline \Pi_{Pij}}

\newcommand{\atf}{\Pi_{Aij}^{(1)}}
\newcommand{\alf}{\Pi_{Aij}^{(0)}}
\newcommand{\ppf}{\Pi_{Pij}}
\newcommand{\pmnf}{\Pi_{Pij}^M}
\newcommand{\Plr} {\Pi_{LR}}
\newcommand{\cond}{\langle\overline q q\rangle}
\newcommand{\condc}{\langle\overline q q\rangle_{\chi}}
\newcommand{\order}{{\cal O}}
\newcommand{\dsp}{\displaystyle}
\newcommand{\be}{\begin{equation}}
\newcommand{\ee}{\end{equation}}
\newcommand{\ba}{\begin{eqnarray}}
\newcommand{\ea}{\end{eqnarray}}
\begin{document}
\begin{titlepage}
\begin{flushright}
CAFPE/15-02\\
LU TP 03-15\\
UGFT/145-02\\
hep-ph/0304222\\
April 2003\\
\end{flushright}
\vfill
\begin{center}
{\Large\bf QCD Short-distance Constraints and
Hadronic Approximations$^\dagger$}
\vfill
{\bf Johan Bijnens$^a$, Elvira G\'amiz$^b$, Edisher Lipartia$^{a*}$
and Joaquim Prades$^b$}\\[0.3cm]
{$^a$Department of Theoretical Physics 2, Lund University,\\
S\"olvegatan 14A, S 223-62 Lund, Sweden}\\[0.5cm]
{$^b$Centro Andaluz de F\'{\i}sica de las 
Part\'{\i}culas Elementales (CAFPE) and \\
Departamento de F\'{\i}sica Te\'orica y
del Cosmos, Universidad de Granada \\Campus de Fuente Nueva, 
E-18002 Granada, Spain}
\end{center}
\vfill
\begin{abstract}
This paper discusses a general class of ladder resummation inspired
hadronic approximations. It is found that this approach
naturally reproduces many successes of single meson per channel
saturation models (e.g. VMD) and NJL based models.
In particular the existence of a constituent quark mass
and a gap equation follows naturally.
We construct an approximation that satisfies a large set of QCD 
short-distance and large $N_c$ constraints and reproduces many hadronic 
observables.

We show how there exists in general a problem between QCD short-distance
constraints for Green Functions and those for form factors and 
cross-sections
following from the quark-counting rule. This problem while expected for
Green functions that do not vanish in purely perturbative QCD also
persists for many Green functions that are order parameters.
\end{abstract}
\vfill

\vfill
\footnoterule
{\footnotesize\noindent$^\dagger$ Supported in part by the European Union RTN
network, Contract No. HPRN-CT-2002-00311  (EURIDICE)}

{\footnotesize\noindent$^*$ On leave of absence from
{\em Laboratory of Information Technologies,
Joint Institute for Nuclear Research, 141980 Dubna, Russia}
and
{\em High Energy Physics Institute, Tbilisi State University, University
St. 9, 380086 Tbilisi, Georgia}.}

\end{titlepage}

\section{Introduction}

Formulating a consistent hadronic approximation to Quantum Chromodynamics (QCD)
is an old and very difficult problem. At low energies the solution to this
problem is Chiral Perturbation Theory (ChPT) but the domain of validity of this
is fairly limited and there tend to be a rather large number of parameters
that needs to be dealt with. It cannot be simply extended to the intermediate
energy domain.
In this paper we describe an
approach based on a few simple assumptions and then try to see
how far this can go. This fits naturally in  the limit
of large number of colours ($N_c$).
In this limit
and assuming confinement, QCD is known
to reduce to a theory of stable hadrons interacting only at tree level
\cite{largeNc}. So the only singularities in amplitudes are produced
by the various tree-level poles occurring. This has long been a problem
for various variants of models incorporating some notion of constituent quarks
like the Nambu--Jona-Lasinio (NJL) models \cite{NJL,NJLreviews,ENJLreview}
or the chiral quark model \cite{CQM}.

The main idea in this paper is to take the underlying
principle of ladder resummation
approaches to hadronic physics and make two successive approximations
in this. First we treat the rungs of the ladder as a type of general
contact interaction and second the remaining loop-integrations that occur,
which are always products of one-loop integrations, we treat
as general everywhere analytic functions. The only singularities that
occur then are those generated by the resummations and we naturally end up
with a hadronic large $N_c$ model.

This is also very close to the treatment
of the (extended) Nambu--Jona-Lasinio models
as given in \cite{BBR,BRZ,BP1} where $n$-point Green
functions\footnote{In the remainder these are
often referred to as $n$-point functions.}
are seen as chains of one-loop bubbles connected by a one-loop
with three or more vertices.
The one-loop bubbles can be seen as one-loop Green functions as well.
The full Green functions there are thus composed of one-loop Green functions
glued together by the (ENJL) couplings $g_V$ and $g_S$. 
One way to incorporate confinement in these ENJL models is by
introducing an infinite number of counterterms to remove all the
unwanted singularities \cite{PPR}. 
In \cite{PPR} it was then argued that
the ENJL approach was basically identical
to a one resonance saturation approach. They then proposed a minimal hadronic
ansatz where one resonance saturation is the underlying principle
and all couplings should be
determined from QCD short-distance and chiral constraints
with the relevant
short-distance constraints those that result from order parameters.
Order parameters are
quantities which would be fully zero if only perturbative QCD without
quark masses and condensates is considered.
This approach has been further discussed for two-point Green functions
in \cite{GP} and applied to some three-point functions in \cite{KN1},
see also the discussions in \cite{Moussallam} for earlier 
similar uses of order parameters.
Problems appear for $n$-point Green functions in that not necessarily
all freedom in the parameters can be fixed by the long-distance chiral 
constraints and/or short-distance
constraints or involve too many unknown constants in the chiral constraints.

In this paper we follow a different scheme.
We {\em assume} that the Green functions are produced by a
ladder-resummation
like ansatz. They consist of bubble-diagrams put together from one-loop
Green functions. We do {\em not} use the (constituent) quark-loop
expressions for these one-loop Green functions but instead consider
them as constants or low-order polynomials in the kinematic variables.
This set of assumptions turns out to be rather constraining in the type
of model that can be constructed. In particular the gap equation for
spontaneous symmetry breaking follows from the requirements of resummation
and the full Ward identities as shown in Section \ref{twopoint}.
The link with constituent quark models is the
fact that given the full Ward identities one can define a constituent
quark mass, obeying a gap equation,
and the one-loop Green functions satisfy the Ward identities
with {\em constituent} quark-masses. In the two-point function
sector this naturally reduces to the approach of \cite{PPR} but it
allows to go beyond two-point functions in a more systematic manner.

In Section \ref{twopoint} we discuss the buildup of the model
and the two-point functions. We first work in the chiral limit and then
add corrections due to current quark masses. Chiral Perturbation
Theory, or low-energy, constraints are naturally satisfied in our approach
which is chiral invariant from the start. Also large $N_c$ constraints are 
satisfied naturally. We show how the short-distance constraints can be 
included.
Section \ref{threepoint} treats several three-point functions and includes
here short-distance constraints coming from form factors and from
the more suppressed combinations of short-distances.

Numerical results are presented in Section \ref{numerics}.
We find a reasonable
agreement for the predictions.

Going beyond the one-resonance saturation
in this approach is difficult as explained in Section \ref{trouble}.
Another point raised is that hadronic models will in general have problems
with QCD short-distance constraints, even if the short-distance behaviour is
an order parameter, we discuss in detail how the
pseudo-scalar--scalar--pseudo-scalar three-point function is a typical
example of this problem in Section \ref{SDProblem}.

We consider this class of models still useful even with the problems
inherent in it. They provide a consistent framework to address the
problems of nonleptonic matrix-elements where in general very many
Green functions with a large number of insertions is needed. The present
approach offers a method to \textsl{analytically} calculate these
Green functions and
thus study the effects of the various ingredients on the final results.
One motivation for this work was to understand many of the rather surprising
features found in the calculations using the ENJL model of the $B_K$ parameter,
the $\Delta I=1/2$ rule, gluonic and electroweak Penguins,
electromagnetic effects and the muon anomalous magnetic
moment\cite{BPx,BPPx} and improve on those calculations.

The Appendix contains expressions for the short-distance properties of
several three-point functions.

\section{Basics of the Model and Two-Point Functions}
\label{twopoint}

\subsection{General}
\label{twopointgeneral}

The Lagrangian for the large $N_c$ ENJL model is
\ba
{\cal L}_{ENJL}
&=&
\sum_{i,j,\alpha}
\overline q_\alpha^i\left\{ \gamma^\mu\left(i\partial_\mu\delta^{ij}+v_\mu^{ij}
+a_\mu^{ij}\gamma_5\right)
-{\cal M}^{ij}-s^{ij}+ip^{ij}\gamma_5\right\}q^j_\alpha
\nonumber\\&&
+2 g_S\sum_{i,j,\alpha,\beta}\left(\overline q^i_{R\alpha} q^j_{L\alpha}\right)
\left(\overline q^j_{L\beta} q^i_{R\beta}\right)
\nonumber\\&&
- g_V\sum_{i,j,\alpha,\beta}
\left(\overline q^i_{L\alpha}\gamma_\mu q^j_{L\alpha}\right)
\left(\overline q^j_{L\beta}\gamma^\mu q^i_{L\beta}\right)
- g_V\sum_{i,j,\alpha,\beta}
\left(\overline q^i_{R\alpha}\gamma_\mu q^j_{R\alpha}\right)
\left(\overline q^j_{R\beta}\gamma^\mu q^i_{R\beta}\right)
\ea
with $i,j$ flavour indices, $\alpha,\beta$ colour indices
and $q_{R(L)} = (1/2)(1+(-)\gamma_5)q$.
The flavour matrices $v,a,s,p$ are external fields and can be used to
generate all the Green functions we will discuss.
The four-quark interactions can be seen as an approximation for the
rungs of a ladder-resummation scheme.

The Green functions generated by functional differentiation w.r.t.
$v^{ij}(x),a^{ij}(x),s^{ij}(x),p^{ij}(x)$ correspond to Green functions of
the currents
\ba
\label{currents}
V_\mu^{ij}(x) &=& \overline q^i_\alpha(x) \gamma_\mu q^j_\alpha(x)\,,
\nonumber\\
A_\mu^{ij}(x) &=& \overline q^i_\alpha(x) \gamma_\mu\gamma_5 q^j_\alpha(x)\,,
\nonumber\\
S^{ij}(x) &=& -\overline q^i_\alpha(x)  q^j_\alpha(x)\,,
\nonumber\\
P^{ij}(x) &=& \overline q^i_\alpha(x) i\gamma_5 q^j_\alpha(x)\,.
\ea
An underlying assumption is that these currents can be
identified with the QCD ones.

In the remainder of this section we will discuss the two-point functions
\ba
\Pi^V_{\mu\nu}(q)^{ijkl}
&=&
i\int d^dx\,e^{i\, q\cdot x}
\langle 0|T\left(V^{ij}_\mu(x)V^{kl}_\nu(0)\right)|0\rangle\,,
\nonumber\\
\Pi^A_{\mu\nu}(q)^{ijkl}
&=&
i\int d^dx\,e^{i\, q\cdot x}
\langle 0|T\left(A^{ij}_\mu(x)A^{kl}_\nu(0)\right)|0\rangle\,,
\nonumber\\
\Pi^S_{\mu}(q)^{ijkl}
&=&
i\int d^dx\,e^{i\, q\cdot x}
\langle 0|T\left(V^{ij}_\mu(x)S^{kl}(0)\right)|0\rangle\,,
\nonumber\\
\Pi^P_{\mu}(q)^{ijkl}
&=&
i\int d^dx\,e^{i\, q\cdot x}
\langle 0|T\left(A^{ij}_\mu(x)P^{kl}(0)\right)|0\rangle\,,
\nonumber\\
\Pi^S(q)^{ijkl}
&=&
i\int d^dx\,e^{i\, q\cdot x}
\langle 0|T\left(S^{ij}(x)S^{kl}(0)\right)|0\rangle\,,
\nonumber\\
\Pi^P(q)^{ijkl}
&=&
i\int d^dx\,e^{i\, q\cdot x}
\langle 0|T\left(P^{ij}(x)P^{kl}(0)\right)|0\rangle\,.
\ea
The other possibilities vanish because of parity. The large $N_c$ limit
requires these to be proportional to $\delta^{il}\delta^{jk}$
and Lorentz and translational invariance allow them to be written in
terms of functions that only depend on $q^2$ and the flavour index
$i,j$.
\ba
\Pi_{\mu\nu}^V(q)_{ijkl} &=&
\left\{\left(q_\mu q_\nu-g_{\mu\nu} q^2\right) \Pi_{Vij}^{(1)}(q^2)
+ q_\mu q_\nu \Pi_{Vij}^{(0)}(q^2)\right\}
\delta^{il}\delta^{jk}\,,
\nonumber\\
\Pi_{\mu\nu}^A(q)_{ijkl} &=&
\left\{\left(q_\mu q_\nu-g_{\mu\nu} q^2\right) \Pi_{Aij}^{(1)}(q^2)
+ q_\mu q_\nu \Pi_{Aij}^{(0)}(q^2)\right\}
\delta^{il}\delta^{jk}\,,
\nonumber\\
\Pi_{\mu}^S(q)_{ijkl} &=&
 q_\mu  \Pi_{Sij}^{M}(q^2)
\delta^{il}\delta^{jk}\,,
\nonumber\\
\Pi_{\mu}^P(q)_{ijkl} &=&
 i q_\mu  \Pi_{Pij}^{M}(q^2)
\delta^{il}\delta^{jk}\,,
\nonumber\\
\Pi^S(q)_{ijkl} &=&
 \Pi_{Sij}(q^2)
\delta^{il}\delta^{jk}\,,
\nonumber\\
\Pi^P(q)_{ijkl} &=&
 \Pi_{Pij}(q^2)
\delta^{il}\delta^{jk}\,.
\ea
These functions satisfy Ward-identities following from chiral symmetry
and the QCD equations of motion
\ba
\label{WI}
q^2 \Pi_{Vij}^{(0)}(q^2) &=& \left(m_i-m_j\right) \Pi_{Sij}^{M}(q^2)\,,
\nonumber\\
q^2 \Pi_{Sij}^{M}(q^2) &=& \left(m_i-m_j\right) \Pi_{Sij}(q^2)
+\cond_i -\cond_j\,,
\nonumber\\
q^2 \Pi_{Aij}^{(0)}(q^2) &=& \left(m_i+m_j\right) \Pi_{Pij}^{M}(q^2)\,,
\nonumber\\
q^2 \Pi_{Pij}^{M}(q^2) &=& \left(m_i+m_j\right) \Pi_{Pij}(q^2)
+\cond_i+\cond_j\,.
\ea
Here we use $\cond_i =
 \sum_\alpha\langle 0|\overline q^i_\alpha q^i_\alpha | 0 \rangle$.

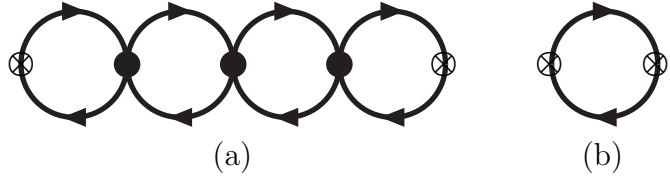
\begin{figure}
\begin{center}
\setlength{\unitlength}{1pt}
\begin{picture}(250,60)(0,-15)
\SetScale{1.}
\SetWidth{2.}
\ArrowArcn(25,25)(20.,0.,180.)
\ArrowArcn(25,25)(20.,180.,0.)
\ArrowArcn(65,25)(20.,0.,180.)
\ArrowArcn(65,25)(20.,180.,0.)
\ArrowArcn(105,25)(20.,0.,180.)
\ArrowArcn(105,25)(20.,180.,0.)
\ArrowArcn(145,25)(20.,0.,180.)
\ArrowArcn(145,25)(20.,180.,0.)
\Text(5,25)[]{{\boldmath$\otimes$}}
\Text(165,25)[]{{\boldmath$\otimes$}}
\Vertex(45,25){5}
\Vertex(85,25){5}
\Vertex(125,25){5}
\Text(85,-10)[]{(a)}
\ArrowArcn(225,25)(20.,0.,180.)
\ArrowArcn(225,25)(20.,180.,0.)
\Text(205,25)[]{{\boldmath$\otimes$}}
\Text(245,25)[]{{\boldmath$\otimes$}}
\Text(225,-10)[]{(b)}
\end{picture}
\end{center}
\caption{\label{figtwopoint}The type of diagrams in large $N_c$ that contribute
to the two-point function. {\boldmath$\otimes$} indicates and insertion
of an external current and $\bullet$ indicates the ENJL four-quark vertex.
(a) the full two-point function. (b) The one-loop two-point function.}
\end{figure}
The type of diagrams that contribute in large $N_c$ to the two-point functions
is depicted in Fig.~\ref{figtwopoint}(a). The contribution from only the
one-loop diagram is depicted in Fig.~\ref{figtwopoint}(b) and we will
generally denote these as $\overline \Pi$.

Under interchange of $i$ and $j$, $\smnf(q^2)$ is anti-symmetric, all others
are symmetric. The one-loop equivalents have the same symmetry properties.

In Refs.~\cite{BRZ,BP1} it was shown that the full two-point functions
can be obtained from the one-loop ones via a resummation procedure
\ba
\label{vssector}
\vtf (q^2)&=& \frac{\vt (q^2)}{1-q^2 g_V \vt(q^2)}\nonumber \\
\vlf (q^2)&=& \frac{1}{\Delta_S(q^2)}[(1-g_S\s (q^2))\vl 
(q^2)+g_S(\SMN(q^2))^2]\nonumber \\
\smnf (q^2)&=& \frac{1}{\Delta_S(q^2)}\SMN(q^2) \nonumber\\
\ssf (q^2) &=& \frac{1}{\Delta_S(q^2)}[(1+q^2 g_V \vl (q^2))\s 
(q^2)-q^2 g_V (\SMN(q^2))^2]\nonumber\\
\Delta_S(q^2)&=&(1+q^2 g_V \vl(q^2) )(1-g_S \s (q^2))+q^2 g_S g_V (\SMN(q^2))^2
\ea
\ba
\label{apsector}
\atf (q^2)&=& \frac{\at (q^2)}{1-q^2 g_V \at(q^2)}\nonumber \\
\alf (q^2)&=& \frac{1}{\Delta_P(q^2)}[(1-g_S\p (q^2))\al 
(q^2)+g_S(\PMN(q^2))^2]\nonumber \\
\pmnf (q^2)&=& \frac{1}{\Delta_P(q^2)}\PMN(q^2) \nonumber\\
\ppf (q^2) &=& \frac{1}{\Delta_P(q^2)}[(1+q^2 g_V \al (q^2))\p 
(q^2)-q^2 g_V (\PMN(q^2))^2]\nonumber\\
\Delta_P(q^2)&=&(1+q^2 g_V \al(q^2) )(1-g_S \p (q^2))+q^2 g_S g_V (\PMN(q^2))^2
\ea
This resummation is only consistent with the Ward Identities, Eq. (\ref{WI}), 
if
the one-loop two-point functions obey the Ward Identities
of Eq.~(\ref{WI}) with the current quark masses $m_i$ replaced
by the constituent quark masses $M_i$ given by
\be
\label{gap}
M_i = m_i -g_S\cond_i\,,
\ee
known as the gap equation. The {\em assumption} of resummation thus leads
to a constituent quark mass picture and one-loop Ward identities
with constituent quark masses.

Using the gap equation and the one-loop Ward identities
the resummation formulas can be simplified
using
\ba
\Delta_S(q^2)&=&1-g_S \s(q^2)+g_V(m_i-m_j)\SMN(q^2)\,,
\nonumber\\
\Delta_P(q^2)&=&1-g_S \p(q^2)+g_V(m_i+m_j)\PMN(q^2)\,.
\ea

Our model assumption is to choose the one-loop functions as basic
parameters rather than have them predicted via the constituent
quark loops. This allows for a theory that has confinement built in
a simple way and at the same time keeps most of the successes of the
ENJL model in low-energy hadronic physics.

We now choose the two-point functions as far as possible as constants
and have thus as parameters in the two-point sector
\be
\label{par2}
\cond_i, g_S, g_V, \PMN,\alt, \SMN,\vlt
\ee
and the remaining one-loop two-point functions can be obtained
from the one-loop Ward identities. As discussed below,
more input will be needed for the three-point functions.
We do not expand higher in momenta in the one-loop two-point functions.
The reason for this is that assuming that $g_V$ and $g_S$ are constants,
expanding the one-loop two-point functions higher in momenta causes
a gap in the large $q^2$ expansion between the leading and the non-leading
terms. Such a gap in powers is not present as we know from perturbative QCD.

\subsection{Chiral Limit}
\label{twopointc}

In the chiral limit, the Ward identity for $\s(q^2)$ becomes singular
and it is better to choose instead as parameters
\be
\label{par2c}
\condc,\Delta,g_S,g_V,\PMNc,\altc,\Gamma,\vltc
\ee
with the parameters $\Delta$, $\Gamma$ defined via
\ba
\label{defDelta}
\cond_i &=& \condc + m_i \Delta + m_i^2 \epsilon + \order(m_i^3)\,,
\nonumber\\
\s(q^2) &=& q^2\Gamma -\frac{\Delta}{1-g_S\Delta} + \order(m_i,m_j)\,.
\ea 

\subsubsection{Short-Distance}
\label{chiralshort}

We define $\Pi_{LR} = \Pi_V - \Pi_A$
and $\Pi_X^{0+1} = \Pi_X^{(0)}+ \Pi_X^{(1)}$ for $X=LR,V,A$ then
the first and third Weinberg sum rules\cite{Weinberg},
\be
\lim_{q^2 \rightarrow -\infty}(q^2\Plr^{(0+1)QCD}(q^2))=0
\quad\mbox{and}\quad
\lim_{q^2 \rightarrow -\infty}(q^4\Plr^{(0)QCD}(q^2))=0\,,
\ee
are automatically satisfied but the second one\,,
\be
\lim_{q^2 \rightarrow -\infty}(q^4\Plr^{(1)QCD}(q^2))=0\,,
\ee
implies the relation
\be
\label{SDcVA}
\altc = \vltc\,.
\ee

Analogs of the Weinberg sum rules exist in scalar-pseudoscalar sector.
With $\Pi_{SP} = \Pi_S-\Pi_P$ we have \cite{Moussallam,RRY}
\be
\lim_{q^2\to-\infty}
\Pi_{SP\,ij}^{QCD}(q^2) = 0\quad\mbox{and}\quad
\lim_{q^2\to-\infty}
(q^2\Pi_{SP\,ij}^{QCD}(q^2)) = 0\,.
\ee
The first one is the equivalent of the first Weinberg sum rule and
is automatically satisfied. The second one implies
\be
\label{SDcSP}
\Gamma = \frac{-\PMNc}{2g_S\condc\left(1-2g_S g_V\condc\PMNc\right)}\,.
\ee

The short-distance relation found in Eq.~(\ref{SDcSP}) does not satisfy
the heat kernel relation for the one-loop two-point functions derived
in \cite{BRZ} in the chiral limit. Note that that heat kernel relation
was the underlying cause of the relation
$m_S = 2 M_q$ between the scalar mass and the constituent quark mass
in ENJL models \cite{BRZ,BP1}.

\subsubsection{Intermediate-Distance}
\label{chiralintermediat}

The two-point functions in the chiral limit can be written as
\ba
\Pi_V^{(1)\chi}(q^2) &=& \frac{2 f_V^2 m_V^2}{m_V^2-q^2}\,,
\nonumber\\
\Pi_A^{(1)\chi}(q^2) &=& \frac{-2 F_0^2}{q^2}+
\frac{\dsp 2 f_A^2 m_A^2}{\dsp m_A^2-q^2}\,,
\nonumber\\
\Pi_P^{M\chi}(q^2) &=& \frac{\dsp 2\condc}{q^2}\,,
\nonumber\\
\Pi_S^\chi(q^2) &=& K_S+\frac{2 F_S^2 m_S^2}{m_S^2-q^2}
\nonumber\\
\Pi_P^\chi(q^2) &=& K_P-\frac{2 F_0^2 B_0^2}{q^2}
\ea

{}From the poles in the two-point functions we can find the various masses.
There is a pole at $q^2=0$ corresponding to the massless pion.
The scalar, vector and axial-vector masses are given by
\ba
m_S^2 &=& \frac{1}{g_S\Gamma\left(1-g_S\Delta\right)}\,,
\nonumber\\
m_V^2 &=& \frac{1}{g_V\vltc}\,,
\nonumber\\
m_A^2 &=& \frac{1-2 g_S g_V \condc \PMNc}{g_V\altc} = 
\left(1-2 g_S g_V \condc \PMNc\right) m_V^2\,.
\ea
The residues at the poles lead to
\ba
2 f_V^2 &=& \vltc\,,
\nonumber\\
2 f_A^2 &=&  \frac{\vltc}{\left(1-2 g_S g_V \condc \PMNc\right)^2}\,,
\nonumber\\
2 F_0^2 &=& \frac{-2 g_S\condc\PMNc}{1-2 g_S g_V \condc \PMNc}\,,
\nonumber\\
K_S &=&K_P = -\frac{1}{g_S}\,,
\nonumber\\
2 F_S^2 &=& \frac{1-g_S\Delta}{g_S}\,,
\nonumber\\
B_0^2 F_0^4 &=& \condc^2
\ea
The short distance constraints lead as expected to
\ba
f_V^2 m_V^2 &=& f_A^2 m_A^2+F_0^2\,,
\nonumber\\
f_V^2 m_V^4 &=& f_A^2 m_A^4\,,
\nonumber\\
K_S &=& K_P\,,
\nonumber\\
F_S^2 m_S^2 &=& F_0^2 B_0^2\,.
\ea

\subsubsection{Long-Distance}
\label{chirallong}

The two-point functions in the chiral limit can be determined from Chiral
Perturbation Theory. This lead to the identification of $B_0$, $F_0$
with the quantities appearing there and in addition
\ba
\label{L10}
L_{10} &=& -\frac{1}{4}\left(f_V^2-f_A^2\right)\,,   \quad
H_1 = -\frac{1}{8}\left(f_V^2+f_A^2\right)\,,
\nonumber\\
32 B_0^2 L_8 &=& 2 F_S^2\,,\quad\quad\quad\quad\quad
 16 B_0^2 H_2 = 2 K_S+ 2 F_S^2
\ea

\subsubsection{Parameters}

Notice that from the six input parameters we can only determine five
from the two-point function inputs.
A possible choice of input parameters is $m_V$, $m_A$, $F_0$, $m_S$ and $F_S$.
The last can be traded for $B_0$ or $\condc$.
The remaining parameter could in principle be fixed from $K_S$ but that
is an unmeasurable quantity.

\subsection{Beyond the Chiral Limit}
\label{twopointbeyond}

The resummation formulas of Sect.~\ref{twopointgeneral} remain valid.
What changes now is that we have values for the current quark masses $m_i$
and corresponding changes in the one-loop functions.
An underlying expectation is that the vertices $g_S$ and $g_V$
are produced by purely gluonic effects and have no light quark-mass dependence.
The first order the quark-mass dependence of $g_V$ and $g_S$
must be zero from short-distance constraints as shown below.

The input parameters are now given by Eq. (\ref{par2}) and we will below
expand them as functions in $m_q$.

\subsubsection{Intermediate-Distance}

The resummation leads to expressions for the two-point functions
which can again be written as one resonance exchange.
\ba 
\label{mesondominance}
\vtf (q^2)&=& -\frac{2\,f_{Sij}^2}{q^2}\,
+\,\frac{2\,f_{Vij}^2 m_{Vij}^2}{m_{Vij}^2-q^2} \,,
\nonumber\\
\vlf(q^2)&=& 2f_{Sij}^2\left(\frac{1}{m_{Sij}^2-q^2}+\frac{1}{q^2}\right) \,,
\nonumber\\
\atf (q^2)&=& -\frac{2\,f_{ij}^2}{q^2}\,
+\,\frac{2\,f_{Aij}^2 m_{Aij}^2}{m_{Aij}^2-q^2}  \,,
\nonumber\\
\alf(q^2)&=& 2f_{ij}^2\left(\frac{1}{m_{ij}^2-q^2}+\frac{1}{q^2}\right) \,,
\nonumber\\
\smnf(q^2)&=&\frac{2F_{Sij}m_{Sij}f_{Sij}}{m_{Sij}^2-q^2} \,,
\nonumber\\
\pmnf(q^2)&=&\frac{2B_{ij}f_{ij}^2}{m_{ij}^2-q^2} \,,
\nonumber\\
\ssf (q^2) &=& K_{Sij}\,+\frac{2\,F_{Sij}^2m_{Sij}^2}{m_{Sij}^2-q^2} \,,
\nonumber\\
\ppf (q^2) &=& K_{Pij}\,+\frac{2\,f_{ij}^2 B_{ij}^2}{m_{ij}^2-q^2} \,.
\ea
These satisfy the Ward Identities (\ref{WI}).
The values of the couplings and masses are given by
\ba \label{LMDparameters}
m_{Vij}^2 &=& \frac{1+g_V\SMN (M_i-M_j)}{g_V \vlt}  \,,
\nonumber\\
m_{Aij}^2 &=& \frac{1+g_V\PMN (M_i+M_j)}{g_V \alt}  \,,
\nonumber\\
m_{Sij}^2 &=& \frac{m_i-m_j}{\SMN}\,\frac{ (1+g_V\SMN (M_i-M_j))}{g_S} \,,
\nonumber\\
m_{ij}^2&=&(m_i+m_j)\frac{1+g_V(M_i+M_j)\PMN}{g_S\PMN} \,,
\nonumber\\
2\,f_{Vij}^2 &=& \frac {\vlt} {(1+g_V\SMN (M_i-M_j))^2} \,,
\nonumber\\
2\,f_{Aij}^2 &=&  \frac {\alt} {(1+g_V\PMN (M_i+M_j))^2} \,,
\nonumber\\
2\,F_{Sij}^2 &=& \frac{M_i-M_j}{g_S\,(m_i-m_j)} \,,
\nonumber\\
2\,f_{Sij}^2 &=& \frac{(M_i-M_j)\SMN}{1+g_V\SMN (M_i-M_j)} \,,
\nonumber\\
2f_{ij}^2&=&\frac{(M_i+M_j)\PMN}{1+g_V\PMN (M_i+M_j)} \,,
\nonumber\\
K_{Sij} &=& K_{Pij} = -\frac{1}{g_S} \,,
\nonumber\\
B_{ij} &=&\frac{1+g_V(M_i+M_j)\PMN}{g_S\PMN} \,.
\ea

\subsubsection{Short-Distance}

In order to proceed we have to expand the input parameters of
Eq. (\ref{par2}) in the quark masses $m_q$.
\ba
\vlt &=& \vltc+(m_i+m_j)\vlti+\order(m_q^2)\,,
\nonumber\\
\alt &=& \altc+(m_i+m_j)\alti+\order(m_q^2)\,,
\nonumber\\
\PMN &=& \PMNc+(m_i+m_j)\PMNi+\order(m_q^2)\,,
\nonumber\\
\s(q^2) &=& q^2\left(\Gamma+(m_i+m_j)\Gamma^I\right)-\frac{\Delta}{1-g_S\Delta}
-\frac{\epsilon}{\left(1-g_S\Delta\right)^2}(m_i+m_j)+\order(m_q^2)\,,
 \ea
The parameters $\epsilon$ and $\Delta$ are defined in the first line of
Eq.~(\ref{defDelta}). The other one-loop two-point functions are derivable
from the one-loop Ward identities.

The chiral limit short-distance constraints Eqs. (\ref{SDcVA})
 and (\ref{SDcSP})
remain valid but there are new constraints on the coefficients of the
quark mass expansions. The derivatives w.r.t. the quark masses of the
two-point functions allow to construct more order parameters
than  $\Pi_{LR}$ and $\Pi_{SP}$. In particular we have\footnote{We have derived
these expressions but they can also be found in \cite{RRY}.}
\ba
\label{SDi2point}
\lim_{q^2\to-\infty}\lim_{m_q\to 0}
\left(q^4\frac{\partial}{\partial m_i}\vtf(q^2)\right) &=& \condc\,,
\nonumber\\
\lim_{q^2\to-\infty}\lim_{m_q\to 0}
\left(q^4\frac{\partial}{\partial m_i}\vlf(q^2)\right) &=& 0\,,
\nonumber\\
\lim_{q^2\to-\infty}\lim_{m_q\to 0}
\left(q^4\frac{\partial}{\partial m_i}\atf(q^2)\right) &=& -\condc\,,
\nonumber\\
\lim_{q^2\to-\infty}\lim_{m_q\to 0}
\left(q^4\frac{\partial}{\partial m_i}\alf(q^2)\right) &=& 2\condc\,,
\nonumber\\
\lim_{q^2\to-\infty}\lim_{m_q\to 0}
\left(q^2\frac{\partial}{\partial m_i}\ssf(q^2)\right) &=&-\frac{3}{2}\condc\,,
\nonumber\\
\lim_{q^2\to-\infty}\lim_{m_q\to 0}
\left(q^2\frac{\partial}{\partial m_i}\ppf(q^2)\right) &=&\frac{1}{2}\condc\,.
\ea
The ones with lower powers of $q^2$ must vanish.
The second and fourth are automatically satisfied
as a consequence from the
Ward identities. $\vlf(q^2)$ only starts at $\order(m_q^2)$
and the $m_i+m_j$ term in $\alf(q^2)$ follows from the Ward identity
and the chiral limit form of $\pmnf$.

The vanishing of those with lower powers of $q^2$ requires that
\be
\lim_{m_q\to0} \frac{\partial g_V}{\partial m_i} =
\lim_{m_q\to0} \frac{\partial g_S}{\partial m_i} = 0\,.
\ee
The first, third, fifth and sixth identities give
\ba
\label{2pointSD}
\vlti &=& g_V^2\left(\vltc\right)^2\condc\,,
\nonumber\\
\alti &=& -g_V^2\left(\vltc\right)^2\condc\,,
\nonumber\\
\Gamma^I &=& -\frac{3}{2} g_S^2\Gamma^2\condc\,,
\nonumber\\
\PMNi &=&  -\frac{1}{4}g_S\left(\PMNc\right)^2
-\frac{1-g_S\Delta}{2 g_S\condc}\, \PMNc
\left(1-4 g_V g_S\condc\PMNc\right)\,.
\ea
This implies that the only new parameter that appears
to include quark masses to first order is $\epsilon$.
The last constraint turns out to be incompatible with short-distance
constraints from three-point functions as discussed below.

\subsubsection{Long-Distance}

The long-distance expansion of our results to $\order(p^4)$
in Chiral Perturbation Theory allows in addition to those already obtained
in the chiral limit also
\be
\label{result1}
L_5 = \frac{1}{16}F_0^6\Big \lbrack \frac{\PMNc(g_S\Delta-1)+2g_S\condc
\PMNi} {(\PMNc)^2g_S^2\condc^3}\Big \rbrack \,.
\ee

\subsubsection{Intermediate-Distance}

The short-distance constraints lead to several relations between resonance
parameters also beyond the chiral limit to first order in current quark masses.
In the vector sector we obtain
\ba
\label{Vrelations}
f_{Vij}^2 m_{Vij}^2 &=& f_{Vkl}^2 m_{Vkl}^2\,,
\nonumber\\
f_{Vij}^2 m_{Vij}^4 - f_{Vkl}^2 m_{Vkl}^4 &=&
-\frac{1}{2}\condc(m_i+m_j-m_k-m_l)\,.
\ea
$V_{ij}$ stands here for the vector degree of freedom built
of quarks with current mass $m_i$ and $m_j$.

The corresponding axial relations are
\ba
\label{Arelations}
f_{Aij}^2 m_{Aij}^2+f_{ij}^2 &=& f_{Akl}^2 m_{Akl}^2+f_{kl}^2\,,
\nonumber\\
f_{Aij}^2 m_{Aij}^4 - f_{Akl}^2 m_{Akl}^4 &=&
\frac{1}{2}\condc(m_i+m_j-m_k-m_l)\,.
\ea

\section{Three-Point Functions}
\label{threepoint}

\begin{figure}
\begin{center}
\setlength{\unitlength}{1pt}
\begin{picture}(100,100)(-10,-10)
\SetScale{1.0}
\SetWidth{2.}
\ArrowLine(5,50)(55,80)
\ArrowLine(55,80)(55,20)
\ArrowLine(55,20)(5,50)
\Text(5,50)[]{{\boldmath$\otimes$}}
\Text(55,80)[]{{\boldmath$\otimes$}}
\Text(55,20)[]{{\boldmath$\otimes$}}
\Text(-5,40)[]{$A, q\rightarrow$}
\Text(55,85)[lb]{$B, p_1\rightarrow$}
\Text(55,10)[lt]{$C, p_2\rightarrow$}
\Text(25,75)[b]{$i$}
\Text(25,30)[t]{$j$}
\Text(60,50)[l]{$k$}
\Text(40,-5)[]{(a)}
\end{picture}
~~~~~~
\setlength{\unitlength}{1pt}
\begin{picture}(100,100)(0,-10)
\SetScale{1.0}
\SetWidth{1.}
\ArrowArcn(12.5,50)(7.5,0.,180.)
\ArrowArcn(12.5,50)(7.5,180.,0.)
\ArrowArcn(27.5,50)(7.5,0.,180.)
\ArrowArcn(27.5,50)(7.5,180.,0.)
\ArrowArcn(42.5,50)(7.5,0.,180.)
\ArrowArcn(42.5,50)(7.5,180.,0.)
\SetWidth{2.}
\ArrowLine(50,50)(90,80)
\ArrowLine(90,80)(90,20)
\ArrowLine(90,20)(50,50)
\SetWidth{1.}
\ArrowArcn(97.5,80)(7.5,0.,180.)
\ArrowArcn(97.5,80)(7.5,180.,0.)
\ArrowArcn(112.5,80)(7.5,0.,180.)
\ArrowArcn(112.5,80)(7.5,180.,0.)
\ArrowArcn(127.5,80)(7.5,0.,180.)
\ArrowArcn(127.5,80)(7.5,180.,0.)
\ArrowArcn(97.5,20)(7.5,0.,180.)
\ArrowArcn(97.5,20)(7.5,180.,0.)
\ArrowArcn(112.5,20)(7.5,0.,180.)
\ArrowArcn(112.5,20)(7.5,180.,0.)
\ArrowArcn(127.5,20)(7.5,0.,180.)
\ArrowArcn(127.5,20)(7.5,180.,0.)
\Text(5,50)[]{{\boldmath$\otimes$}}
\Text(135,80)[]{{\boldmath$\otimes$}}
\Text(135,20)[]{{\boldmath$\otimes$}}
\Text(70,-5)[]{(b)}
\end{picture}
\end{center}
\caption{\label{figthreepoint} The $\Pi^+$ contribution to a generic
three-point function. (a) The flavour and momentum flow indicated on a one-loop
diagram. (b) A generic large $N_c$ diagram with the resummation in terms
of bubbles. Note that the resummation leads to full two-point functions.}
\end{figure}
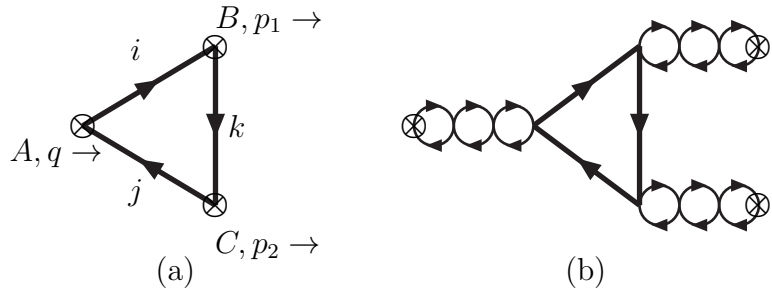

A generic three-point function of currents $A,B,C$ chosen from
the currents in Eq. (\ref{currents}) is defined as
\be
\Pi^{ABC}(p_1,p_2)^{ijklmn}
= i^2\int d^dx d^dy \,e^{ip_1\cdot x}e^{i p_2\cdot y}
\langle 0 | T\left(A^{ij}(0)B^{kl}(x)C^{mn}(y)\right)|0\rangle\,.
\ee
In the large $N_c$ limit these can only have two types of
flavour flow
\be
\label{defpm}
\Pi^{ABC}(p_1,p_2)^{ijklmn}
=
\Pi^{ABC+}(p_1,p_2)^{ijk}\delta^{il}\delta^{jm}\delta^{kn}
+\Pi^{ABC-}(p_1,p_2)^{ijl}\delta^{in}\delta^{jk}\delta^{lm}
\ee
and they satisfy
\be
\Pi^{ABC-}(p_1,p_2)^{ijl}
=
\Pi^{ACB+}(p_2,p_1)^{ijl}\,.
\ee
The flavour and momentum flow of $\Pi^{ABC+}(p_1,p_2)^{ijk}$
is indicated in Fig.~\ref{figthreepoint}{(a)}. In the remainder we
will always talk about the $\Pi^+$ part only but drop the superscript $+$.
We also use $q = p_1+p_2$.
A generic contribution to the three-point function is shown in
Fig.~\ref{figthreepoint}(b). The internal vertices are given
by $g_V$ and $g_S$. In Ref.~\cite{BP1} it was shown on two examples
how this resummation can be performed for some three-point functions.
Many other cases were worked out for the work on non-leptonic
matrix-elements in Refs.\cite{BPx,BPPx}.

Here we will make the assumption of resummation for the
three-point functions just as we did for the two-point
functions in Sect.~\ref{twopoint}. It can again be shown that the Ward
identities for the full three-point functions and the resummation
together require that the one-loop
three-point functions satisfy the one-loop Ward identities with the
constituent masses given by the gap equation (\ref{gap}).

We will once more assume that the three-point functions are
constants or low-order polynomials of the kinematical variables, in 
agreement with the large $N_c$ limit Green functions structure. 
It turns out that the combination of one-loop Ward identities
and short distance constraints is very powerful in restricting the
number of new free parameters appearing in the three-point functions.
This could already be seen in Sect.~\ref{twopointbeyond}, since the
derivative w.r.t. a quark mass of a two-point function is a three-point
function with one of the momenta equal to zero.

A full analysis of three-point functions is in progress. Here we give
a few representative examples.

\subsection{The Pseudoscalar-Scalar-Pseudoscalar Three-Point Function
and the Scalar Form Factor}
\label{sectPSP}

The Pseudoscalar-Scalar-Pseudoscalar three-point function can be calculated
from the class of diagrams depicted in Fig.~\ref{figthreepoint}(b)
using the methods of \cite{BP1} and reads for the case of $m_i=m_k$
\ba
\label{PSPp}
\Pi^{PSP}(p_1,p_2)^{ijk}
& \equiv &\Big \lbrace 1+g_S\Pi_S(p_1^2)_{ki} \Big \rbrace
\nonumber\\
&&\times \Big \lbrace \overline \Pi^{PSP}(p_1,p_2)^{ijk}\,
  (1+g_S\Pi_P(q^2)_{ji})\,
  (1+g_S\Pi_P(p_2^2)_{jk})\nonumber\\
&&+\overline\Pi^{ASP}_{\mu}(p_1,p_2)^{ijk}\,(-g_V i q^{\mu}\Pi^M _P(q^2)_{ji})
(1+g_S\Pi_P(p_2^2)_{jk})\nonumber\\
&&+\overline\Pi^{PSA}_{\nu}(p_1,p_2)^{ijk}\, (1+g_S\Pi_P(q^2)_{ji})
(g_V i p_2^{\nu}\Pi^M _P(p_2^2)_{jk})\nonumber\\
&&+\overline\Pi^{ASA}_{\mu \nu}(p_1,p_2)^{ijk}\,
(-g_V i q^{\mu}\Pi^M _P(q^2)_{ji})
(g_V i p_2^{\nu}\Pi^M _P(p_2^2)_{jk})\Big \rbrace \, .
\ea
The general case has also terms involving one-loop three-point functions
with a vector ($V$) instead of a scalar ($S$).
The one-loop Ward identities can be used to rewrite $\overline\Pi^{ASP}$,
$\overline\Pi^{PSA}$ and $\overline\Pi^{ASA}$ in terms of
$\overline\Pi^{PSP}$ and one-loop two-point functions.

The one-loop three-point function  $\overline\Pi^{PSP}$ is in turn fully fixed
by the one-loop Ward Identities. Let us illustrate the derivation,
one Ward Identity is
\be
i p_2^\mu \overline\Pi^{PSA}_\mu(p_1,p_2)^{ijk} =
-(M_j+M_k) \overline\Pi^{PSP}(p_1,p_2)^{ijk}
+\overline\Pi_{Sik}(p_1^2)-\overline\Pi_{Pij}(q^2)\,.
\ee
Putting $p_1^2=p_2^2=q^2=0$ this determines
\be
\overline\Pi^{PSP}(0,0)^{ijk}
= \frac{1}{M_j+M_k}\left\{
\frac{\cond_k-\cond_i}{M_i-M_k}+\frac{\cond_i+\cond_j}{M_i+M_j}
\right\}
\,.
\ee
The same result follows from the identities for
$q^\mu\overline\Pi^{ASP}_\mu(p_1,p_2)^{ijk}$ and 
$p_1^\mu\overline\Pi^{PVP}_\mu(p_1,p_2)^{ijk}$.

The next term, linear in $q^2, p_1^2, p_2^2$, can be derived as well,
since the relevant combinations of the three-point functions with
one vector or axial-vector
can be determined from Ward identities involving three-point
functions with two vector or axial-vector currents.

We only quote here the chiral limit result
\ba
\label{PSPc}
\overline\Pi^{PSP}(p_1,p_2)^{\chi}
&=&\frac{1}{2g_S^2\left(1-g_S\Delta\right)\condc}
-\frac{p_1^2}{8g_S\condc}\left(4\Gamma- \frac{2\PMNi}{\left(1-g_S\Delta\right)}
+\frac{\PMNc}{g_S\condc}
\right)
\nonumber\\&&
-\frac{q^2+p_2^2}{8 g_S\condc}
\left( \frac{2\PMNi}{\left(1-g_S\Delta\right)}+\frac{\PMNc}{g_S\condc}\right)
\,.
\ea

{}From the $q^2$ dependence of the full Green-function at low energies
we can also derive $L_5$, the result agrees with Eq.~(\ref{result1}) as
it should.

We can look at two different types of short-distance constraints.
First, using the methods of exclusive processes
in perturbative QCD \cite{BrodskyLepage}, it can be shown that the scalar
form factor in the chiral limit should decrease as $1/p_1^2$. 
Phenomenologically, this short-distance behaviour has been also imposed in 
\cite{JOP01} to calculate the scalar form factor. It was checked that 
this behaviour agrees with data. 
Using the LSZ reduction formulas the scalar form factor
of the pion in the chiral limit is
\be
F^\chi_S(p_1^2) = \lim_{q^2,p_2^2\to 0}
\frac{q^2 p_2^2}{-2 F_0^2 B_0^2} \Pi^{PSP}(p_1,p_2)^{\chi}
\ee
and it can be written in a simpler form\footnote{Notice that in order
to have the usual
scalar form factor we need to add the
$\Pi^+$ and $\Pi^-$ of Eq.~(\ref{defpm}). The formulas here refer only
to $\Pi^+$.}
\be
F^\chi_S(p_1^2) = B_0 \frac{m_S^2}{m_S^2-p_1^2}
\left(1+p_1^2 \left(\frac{4 L_5}{F_0^2}-\frac{1}{m_S^2}\right)\right)\,.
\ee
The short-distance requirement on $F_S^\chi(p_1^2)$
thus requires $L_5$ to have its
resonance dominated value
\be
L_5 = \frac{F_0^2}{4 m_S^2}\,.
\ee
This gives a new relation between the input parameters, after using
Eq.~(\ref{SDcSP}),
\be
\label{3pointSD}
\PMNi = \frac{(1-g_S \Delta)\PMNc}{2 g_S\condc}
\left(-1+4g_V g_S \condc\PMNc\right)\,.
\ee
This constraint is not compatible with Eq.~(\ref{2pointSD}).

The three-point function $\Pi^{PSP}(p_1,p_2)^{ijk}$ is an order parameter
in the sense described above. Its short-distance properties can thus be used
to constrain the theory. The short-distance behaviour is
\be
\label{SD_PSP}
\lim_{\lambda\to \infty} \Pi^{PSP}(\lambda p_1,\lambda p_2)^{\chi}
= 0\,.
\ee
This is automatically satisfied by our expression (\ref{PSPcf}).

The entire $\Pi^{PSP\chi}$ can be written in a simple fashion
\be
\label{PSPcf}
\Pi^{PSP}( p_1, p_2)^{\chi}
= -2 F_0^2 B_0^3\frac{m_S^2}{q^2 p_2^2 (m_S^2-p_1^2)}
\left(1 + b (q^2 + p_2^2-p_1^2)\right)
\ee
with
\be
 b = 0 (\mbox{~Eq.~(\ref{3pointSD})})
\quad\mbox{or}\quad
 b = \frac{F_0^4}{8\condc^2} = \frac{1}{8 B_0^2}
  (\mbox{~Eq.~(\ref{2pointSD})})\,.
\ee
The short distance relation $\lim_{\lambda\to\infty}F_S^\chi(\lambda p_1^2)
=0$ has no $\alpha_S$ corrections. We therefore consider the constraint
Eq.~(\ref{3pointSD}) to be more reliable than the one from Eq.~(\ref{2pointSD}).

\subsection{The Vector-Pseudoscalar-Pseudoscalar Three-Point Function
and the Vector Form Factor}
\label{sectVPP}

We can repeat the analysis of Sect.~\ref{sectPSP} now for the $VPP$
three-point function. The results will be very similar to there and apply
to the vector (electromagnetic) form factor.
We keep here to the simpler case of $m_i=m_j$.
The resummation leads to\cite{BP1}

\ba 
\label{VPPgeneral}
\Pi^{VPP}_{\mu } (p_1,p_2)^{ijk} & = &
\Big \lbrace g^{\mu \nu}
-g_V\Pi^{V}_{\mu \nu}(q)^{ij}\Big \rbrace \nonumber\\
&& \times \Big \lbrace \overline \Pi^{VPP}_{\nu}(p_1,p_2)^{ijk}
\Big(1+g_S \Pi_{Pik}(p_1^2)\Big)\Big(1+g_S \Pi_{Pkj}(p_2^2)\Big)\nonumber\\
&& +\overline\Pi_{\nu\beta}^{VPA}(p_1,p_2)^{ijk}
   \Big(1+g_S \Pi_{Pik}(p_1^2)\Big)
  \Big(g_V\,i\,p_2^{\beta} \,\Pi_{Pkj}^M(p_2^2)\Big)\nonumber\\
&&+\overline\Pi_{\nu\alpha}^{VAP}(p_1,p_2)^{ijk}
\Big(g_V\,i\,p_1^{\alpha} \,\Pi_{Pik}^{M} (p_1^2)\Big)\,
\Big(1+g_S \Pi_{Pkj}(p_2^2)\Big)\nonumber\\
&&+\overline \Pi_{\nu\alpha\beta}^{VAA}(p_1,p_2)^{ijk}\,
\Big(g_V\,i\,p_1^{\alpha} \,\Pi_{Pik}^M (p_1^2)\Big)\,
\Big(g_V\,i\,p_2^{\beta} \,\Pi_{Pkj}^M (p_2^2)\Big)\Big\rbrace\,.
\ea
We can again use the Ward Identities to rewrite this in terms of
two-point functions and $\overline\Pi^{VPP}_{\mu}(p_1,p_2)^{ijk}$
only. 

We now expand in $p_1^2, p_2^2$
and $(p_1+p_2)^2=q^2$.
\be
\label{VPPexpansion}
\overline\Pi^{VPP}_{\mu}(p_1,p_2)^{ijk}
= p_{1\mu} \overline\Pi^{VPPijk}_1
+ p_{2\mu}\overline\Pi^{VPPijk}_2
+ C^{VPP}_{ijk} \left(q\cdot p_2\, p_{1\mu}-q\cdot p_1\, p_{2\mu}\right)\,.
\ee
The one-loop WI imply
\ba
\overline\Pi^{VPPijk}_1 &=&
  \frac{-\SMN+\overline\Pi_{Pik}^M}{M_j+M_k}
\nonumber\\
\overline\Pi^{VPPijk}_2 &=&
\frac{-\SMN-\overline\Pi_{Pjk}^M}{M_i+M_k}\,.
\ea
The next term in the expansion depends only on one constant. This follows
from the assumption (in the previous subsection) that $\overline\Pi^{SPP}$
contains no terms more than linear in $p_1^2,p_2^2,q^2$. The form
given in Eq.~(\ref{VPPexpansion}) includes this assumption already.
This extra constant can be determined from the fact that the pion vector
factor should decrease as $1/q^2$ for large $q^2$. Extracting the chiral
limit\footnote{This argument is also valid outside the chiral limit.}
vector form factor
via
\be
F^\chi_V(q^2) = \lim_{p_1^2,p_2^2\to0}\frac{p_1^2 p_2^2}{2F_0^2B_0^2}
\Pi^{VPP}_1(p_1,p_2)^\chi\,.
\ee
The subscript one means the coefficient of $p_{1\mu}$ in the
expansion
\be
\Pi_\mu^{VPP}(p_1,p_2)= p_{1\mu}\Pi^{VPP}_1(p_1,p_2)
+p_{2\mu}\Pi^{VPP}_2(p_1,p_2).
\ee
The short-distance requirement then determines
\be
\label{valueCVPP}
C_{\chi}^{VPP} = \left(\PMNc\right)^2 g_V^2\vltc.
\ee
The ChPT expression for the pion vector form factor yields then
\be
\label{L9}
L_9 = \frac{F_0^2}{2} g_V \vltc = \frac{1}{2}\frac{F_0^2}{m_V^2}\,.
\ee

The full chiral limit three-point function can be written in a simple fashion
\be
\label{VPPcf}
\Pi_{\mu}^{VPP}(p_1,p_2)^\chi = \frac{-2 F_0^2 B_0^2}{p_1^2 p_2^2}
\frac{m_V^2}{m_V^2-q^2} 
\left(p_{1\mu}-p_{2\mu}+ A (p_2^2 - p_1^2 )(p_{1\mu}+p_{2\mu})\right)\,,
\ee
with
\be
A = g_V \vltc = \frac{1}{m_V^2}\,.
\ee

\subsection{The Scalar-Vector-Vector Three-Point function}
\label{SVV}

The Scalar-Vector-Vector three-point function has been used
to discuss the properties of the scalars in Ref.~\cite{Moussallam}.
The relation between the full and the one-loop functions in the case of
all masses equal is
\ba
\Pi^{SVV}_{\mu \nu}(p_1,p_2)^{iii}& \equiv &
 \lbrace g_{\mu \alpha}-g_V\Pi^V_{\mu\alpha}(p_1)^{iiii}\rbrace \times 
\lbrace g_{\nu \beta}-g_V\Pi^{V}_{\nu\beta}(p_2)^{iiii}\rbrace
\nonumber\\
&&\times \Big \lbrace 1+g_S\Pi_{Sii}(q^2) \Big \rbrace
\overline\Pi^{SVV}_{\alpha \beta}(p_1,p_2)^{iii}\,.
\ea
In the equal mass case both the full and the one-loop three-point function are
fully transverse.

The one-loop two-point functions expanded to second order in the momenta
is fully determined from the Ward Identities via
\ba
\overline\Pi^{SVV}_{\mu\nu}(p_1,p_2)^{ijk} &=&
\overline\Pi^{SVVijk}_1 g_{\mu\nu}
+\overline\Pi^{SVVijk}_2 \left(p_{2\mu} p_{1\nu}-p_1\cdot p_2\,g_{\mu\nu}
\right)\,,
\nonumber\\
\overline\Pi^{SVVijk}_1 &=&
\frac{1}{M_j-M_i}\left\{\left(M_i-M_k\right)\overline\Pi^M_{Sik}
-\left(M_j-M_k\right)\overline\Pi^M_{Sjk}\right\}\,,
\nonumber\\
\overline\Pi^{SVVijk}_2 &=&
\frac{\overline\Pi^{(0+1)}_{Vik}-\overline\Pi^{(0+1)}_{Vjk}}{M_j-M_i}\,.
\ea
In the chiral limit these expressions reduce to
\ba
\nonumber\\
\overline\Pi^{SVV\chi}_1 &=&
0\,,
\nonumber\\
\overline\Pi^{SVV\chi}_2 &=&
-\frac{\vlti}{1-g_S\Delta}\,.
\ea
The expression for the chiral limit full three-point functions is very simple
\be
\label{SVVcf}
\Pi^{SVV}_{\mu\nu}(p_1,p_2)^\chi = A \frac{m_S^2}{m_S^2-q^2}
\,\frac{m_V^2}{m_V^2-p_1^2}
\,\frac{m_V^2}{m_V^2-p_2^2}\left(p_{2\mu} p_{1\nu}-p_1\cdot p_2\, g_{\mu\nu}
\right)\,.
\ee
with
\be
A = -\vlti = -\frac{\condc}{m_V^4}\,.
\ee
This also satisfies the QCD short-distance requirement
\be
\lim_{\lambda\to\infty}\Pi^{SVV}_{\mu\nu}(\lambda p_1, \lambda p_2)^\chi
= 0\,.
\ee

\subsection{The Pseudoscalar--Vector--Axial-vector Three-Point Function}
\label{PVA}

This three-point functions has been studied in a related way
in Ref.~\cite{KN1}. The expression for the full
Pseudoscalar--Vector--Axial-vector three-point function in terms of the
one-loop one and two-point functions is in the case of $m_i = m_k$:
\ba
\label{PVAf}
\Pi^{PVA}_{\mu \nu} (p_1,p_ 2)^{ijk} & = &
\lbrace g^{\mu \beta}-g_V\Pi_{V\,ik}^{\mu \beta}(p_1)\rbrace
 \nonumber\\
&& \times \Bigg \lbrace 
(1+g_S \Pi_{Pij}(q^2))
(g^{\nu \gamma}-g_V\Pi^{A\nu \gamma}(p_2)^{kj})
\overline\Pi^{PVA}_{\beta \gamma}(p_1,p_2)^{ijk}
\nonumber\\
&& + g_V\,\Pi_{Pij}^M(q^2)(g^{\nu \gamma}
-g_V\Pi^{A\nu \gamma}(p_2)^{kj})
\nonumber\\
&&\times \Big(-(M_i+M_j)\,
   \overline\Pi^{PVA}_{\beta \gamma}(p_1,p_2)^{ijk}
  -i\,\overline\Pi^{V}_{\beta\gamma}(p_1)^{ik}
  +i\,\overline\Pi^{A}_{\beta\gamma}(p_2)^{jk}\Big)
\nonumber\\
&&+(1+g_S\Pi_{Pij}(q^2))g_S\,i\,p_2^{\nu}\Pi_{Pkj}^M(p_2^2)
 \overline\Pi^{PVP}_{\beta}(p_1,p_2)^{ijk}
\nonumber\\
&&+g_S\,g_V\, i \,p_2^{\nu}\Pi_{Pij}^M(q^2) \Pi_{Pkj}^M(p_2^2)
\nonumber\\
&&\times \Big ( -(M_i+M_j)\,\overline\Pi^{PVP}_{\beta}(p_1,p_2)^{ijk}
+\overline\Pi^S_{\beta}(p_1)^{ik}-i\,\overline\Pi^P_{\beta}(p_2)_{jk}\Big)
\Bigg \rbrace\,,
\ea
where we have used the Ward identities
\ba
-i\,q^{\alpha}\,\overline\Pi^{AVA}_{\alpha \beta \gamma}(p_1,p_2)^{ijk}
 & = & -(M_i+M_j)\,\overline\Pi^{PVA}_{\beta \gamma}(p_1,p_2)^{ijk}
-i\,\overline\Pi^{V}_{\beta\gamma}(p_1)^{ik}
+i\,\overline\Pi^{A}_{\beta\gamma}(p_2)^{jk}\,,
\nonumber\\
-i\,q^{\alpha}\,\overline\Pi^{AVP}_{\alpha \beta}(p_1,p_2)^{ijk} & = & 
  -(M_i+M_j)\,\overline\Pi^{PVP}_{\beta}(p_1,p_2)^{ijk}
+\overline\Pi^S_{\beta}(p_1)^{ik}
-i\,\overline\Pi^P_{\beta}(p_2)^{jk}\,.
\ea

The one-loop three-point function up to second order in the momenta is
determined fully from the one-loop Ward Identities.
\ba
\overline\Pi^{PVA}_{\mu\nu}(p_1,p_2)^{ijk} &=&
\overline\Pi^{PVAijk}_1 \, g_{\mu\nu}
+ \overline\Pi^{PVAijk}_2\left(p_{2\mu}p_{1\nu}-p_1\cdot p_2\,g_{\mu\nu}\right)
\nonumber\\&&
+ C^{PVA}_{ijk}
\left(q\cdot p_1\,g_{\mu\nu}-p_{1\mu}p_{1\nu}-p_{2\mu} p_{1\nu}
\right)
\ea
with
\ba
\overline\Pi^{PVAijk}_1 &=&\frac{i}{M_i+M_j}
\left\lbrace\left(M_j+M_k\right)\,\overline\Pi^M_{Pjk}
-\left(M_i-M_k\right)\,\overline\Pi^M_{Sik}\right\rbrace\,,
\nonumber\\
\overline\Pi^{PVAijk}_2 &=&\frac{i}{M_i+M_j}
\left(\overline\Pi^{(0+1)}_{Vik}-\overline\Pi^{(0+1)}_{Ajk}\right)\,,
\nonumber\\
C^{PVA}_{ijk} &=& i (M_j+M_k) C^{VPP}_{kij}\,.
\ea

This expression can be worked out in the chiral limit using the values
obtained earlier and compared with the chiral limit ChPT 
expression for this amplitude
(see e.g. Ref.~\cite{KN1}).
\ba
\Pi_{PVA}^{\mu\nu\,ChPT}&=& 2\,i\condc \Big\lbrace\Big\lbrack
\frac{(p_1+2p_2)^{\mu}p_2^{\nu}}{p_2^2q^2}-\frac{g^{\mu\nu}}{q^2}\Big\rbrack
\nonumber\\&&
+(p_2^{\mu}p_1^{\nu}-\frac{1}{2}(q^2-p_1^2-p_2^2)g^{\mu\nu}) 
\frac {4}{F_0^2q^2}(L_9+L_{10})\nonumber\\&&
+(p_1^2p_2^{\mu}p_2^{\nu}+p_2^2p_1^{\mu}p_1^{\nu}-\frac{1}{2}(q^2-p_1^2-p_2^2)
p_1^{\mu}p_2^{\nu}-p_1^2p_2^2g^{\mu\nu})\frac{4}{F_0^2p_2^2q^2}L_9
\ea
and leads to values of $L_9$ compatible with those obtained
in Eq.~(\ref{L9}) and
\be
L_{10} = -\frac{1}{2} F_0^4g_V \vltc \frac{(g_V g_S\condc\PMNc-1)} 
{g_S \condc \PMNc}\,,
\ee
which is the same  as Eq.~(\ref{L10}).

The three-point function in the chiral limit has a simple
expression of the form
\ba
\label{PVAc}
\Pi^{PVA}_{\mu\nu}(p_1,p_2)^\chi
&=& 
-\frac{2i\condc}{\left(p_1^2-m_V^2\right)q^2}
\left\lbrace\frac{P_{\mu\nu}(p_1,p_2)\,(m_A^2-m_V^2)
+Q_{\mu\nu}(p_1,p_2)}{p_2^2-m_A^2}
-\frac{2 Q_{\mu\nu}(p_1,p_2)}{p_2^2}
\right\rbrace
\nonumber\\&&+\frac{-2 i\condc}{p_2^2 q^2}
\left(p_{1\mu}p_{2\nu}+2p_{2\mu}p_{2\nu}-p_2^2 g_{\mu\nu}\right)\,.
\ea
The tensors $P_{\mu\nu}$ and $Q_{\mu\nu}$ are transverse and defined by 
\ba
P_{\mu\nu} (p_1,p_2) &=&
 p_{2\mu}p_{1\nu}-p_1 \cdot p_2\,g_{\mu\nu}
\nonumber\\
Q_{\mu\nu} (p_1,p_2) &=&
p_1^2\,p_{2\mu}p_{2\nu}+p_2^2\,p_{1\mu}p_{1\nu}-
p_1\cdot p_2\, p_{1\mu} p_{2\nu}-p_1^2 p_2^2\,g_{\mu\nu}\,.
\ea

By construction, this function satisfies the chiral Ward identities 
(see e.g. \cite{KN1})
\ba
p_{1 \mu}\,\Pi_{PVA}^{\mu \nu }(p_1,p_2) &=& -2\,i \, \condc 
\Big \lbrack \frac {p_2^{\nu}}{p_2^2}- \frac{q^{\nu}}{q^2}\Big \rbrack 
\nonumber\\
p_{2\nu}\,\Pi_{PVA}^{\mu \nu }(p_1,p_2) &=& -2\,i \, \condc 
\frac {q^{\mu}}{q^2}
\ea
that are the same as those involving the one-loop function 
$\bar \Pi_{PVA}^{\mu \nu}$ but replacing the constituent masses by current 
quark masses.
The QCD short-distance relation
\be
\lim_{\lambda\to\infty}\Pi^{PVA}_{\mu\nu}(\lambda p_1,\lambda p_2)^\chi = 0\,.
\ee
is also obeyed.

\subsection{The Pseudo-scalar--Axial-vector--Scalar Three-Point Function}
\label{PAS}

Another order parameter is the sum of the Pseudoscalar--Axial-vector--Scalar
and Scalar--Axial-vector--Pseudoscalar three-point functions. 
These functions can be written in terms of the corresponding 
one-loop functions and the two-point functions following the same method 
as in the other sections

For the simpler case $m_j=m_k$
\ba
\label{PASf}
\Pi^{PAS}_{\mu} (p_1,p_2)^{ijk} & = & 
\lbrace 1 + g_S \Pi_{Sjk}(p_2^2)\rbrace \nonumber\\
&& \times \Big \lbrace 
\overline\Pi^{PAS\gamma}(p_1,p_2)^{ijk}
\Big(1+g_S \Pi_{Pij}(q^2)\Big) 
\Big(g_{\mu \gamma}-g_V\Pi^{A}_{\mu \gamma}(p_1)^{ki}\Big)
\nonumber\\
&& +\overline\Pi^{AAS\alpha\gamma}(p_1,p_2)^{ijk}
\Big(-g_V\,i\,q_{\alpha} \,\Pi_P^{Mij} (q^2)\Big)
\Big(g_{\mu \gamma}-g_V\Pi^{A}_{\mu \gamma}(p_1)^{ki}\Big)
\nonumber\\
&& +\overline\Pi^{PPS}(p_1,p_2)^{ijk}
\Big(1+g_S \Pi_{Pij}(q^2)\Big) 
\Big(g_S \,i\,p_{1\mu}\,\Pi_{Pki}^M (p_1^2)\Big)
\nonumber\\
&& +\overline\Pi^{APS}_{\alpha}(p_1,p_2)^{ijk}
\Big(-g_V\,i\,q^{\alpha} \,\Pi_P^{Mij} (q^2)\Big)
\Big(g_S \,i\,p_{1\mu}\,\Pi_{Pki}^M (p_1^2)\Big) \Big\rbrace
\ea
and for the case $m_i=m_j$
\ba
\label{SAPf}
\Pi^{SAP}_{\mu} (p_1,p_2)^{ijk} & = & 
\lbrace 1 + g_S \Pi_{Sij}(q^2)\rbrace \nonumber\\
&& \times \Big \lbrace 
\overline\Pi^{SAP\gamma}(p_1,p_2)^{ijk}
\Big(1+g_S \Pi_{Pjk}(p_2^2)\Big) 
\Big(g_{\mu \gamma}-g_V\Pi^{A}_{\mu \gamma}(p_1)^{ki}\Big)
\nonumber\\
&& +\overline\Pi^{SAA{\alpha\gamma}}(p_1,p_2)^{ijk}
\Big(g_V\,i\,p_{2\alpha} \,\Pi_P^{Mij} (p_2^2)\Big)
\Big(g_{\mu \gamma}-g_V\Pi^{A}_{\mu \gamma}(p_1)^{ki}\Big)
\nonumber\\
&& +\overline\Pi^{SPP}(p_1,p_2)^{ijk}
\Big(1+g_S \Pi_{Pjk}(p_2^2)\Big) 
\Big(g_S \,i\,p_{1\mu}\,\Pi_{Pki}^M (p_1^2)\Big)
\nonumber\\
&& +\overline\Pi^{SPA\alpha}(p_1,p_2)^{ijk}
\Big(g_V\,i\,p_{2\alpha} \,\Pi_P^{Mjk} (p_2^2)\Big)
\Big(g_S \,i\,p_{1\mu}\,\Pi_{Pki}^M (p_1^2)\Big)
\ea

The most general expressions for the one-loop three-point functions 
$\overline\Pi^{SAP}_{\gamma}(p_1,p_2)^{ijk}$ and 
$\overline\Pi^{SAP}_{\gamma}(p_1,p_2)^{ijk}$ up to order $O(p^3)$ and 
compatible with all the symmetries
\be
\label{PASexpansion}
\overline\Pi^{PAS}_{\mu}(p_1,p_2)^{ijk}
= p_{1\mu} \overline\Pi^{PASijk}_1
+ p_{2\mu}\overline\Pi^{PASijk}_2
+ C^{PAS}_{ijk}\,\left(p_1^2\,p_{2\mu}
-p_1\cdot p_2\, p_{1\mu}\right)\,
\ee
\be
\label{SAPexpansion}
\overline\Pi^{SAP}_{\mu}(p_1,p_2)^{ijk}
= p_{1\mu} \overline\Pi^{SAPijk}_1
+ p_{2\mu}\overline\Pi^{SAPijk}_2
- C^{PAS}_{kji} \left(p_1^2\,p_{2\mu}
-p_1\cdot p_2\, p_{1\mu}\right)\,
\ee

There is only one constant at order $O(p^3)$ that remains unknown when we 
apply all the symmetry criteria. The functions in the term of order 
$O(p)$ are fully determined by the use of the one-loop Ward identities
 
\ba
\overline\Pi^{PASijk}_1 &=& 
i\,\frac{\overline\Pi_{Pij}^M -\overline\Pi_{Pik}^M }{M_j-M_k}
\nonumber\\
\overline\Pi^{PASijk}_2 &=&
i\,\frac{\overline\Pi_{Sjk}^M +\overline\Pi_{Pik}^M }{M_i+M_j}
+i\,\frac{\overline\Pi_{Pij}^M -\overline\Pi_{Pik}^M }{M_j-M_k}
\nonumber\\
\overline\Pi^{SAPijk}_1 &=& 
i\,\frac{\overline\Pi_{Sij}^M -\overline\Pi_{Pik}^M }{M_j+M_k}
\nonumber\\
\overline\Pi^{SAPijk}_2 &=& 
i\,\frac{\overline\Pi_{Sij}^M -\overline\Pi_{Pik}^M }{M_j+M_k}
+i\,\frac{\overline\Pi_{Pjk}^M -\overline\Pi_{Pik}^M }{M_i-M_j}
\ea

Using the values of the coupling constants $L_5$ and $L_8$ we obtained 
from two-point functions, the functions $\Pi^{PAS}_{\mu} (p_1,p_ 2)^{ijk}$ 
and $\Pi^{SAP}_{\mu} (p_1,p_ 2)^{ijk}$ have the correct behaviour at 
long distance as described by Chiral Perturbation Theory. In this limit 
the unknown constant $C^{PAS}_{ijk}$ is not involved. 

The sum of the two three-point functions in the chiral limit
can be written in a fairly simple
fashion
\ba
\Pi^{PAS+SAP}_{\mu} (p_1,p_ 2)^{\chi} &=& i B_0^2 F_0^2
\frac{m_S^2}{(m_S^2-q^2)(m_S^2-p_2^2)(m_A^2-p_1^2)p_2^2q^2p_1^2}\nonumber\\
&&\times\Bigg\{ p_{2\mu}\,4\,
(m_A^2 +  D^{PAS}_{\chi}p_1^2)p_1^2(q^2-p_2^2)\nonumber\\
&&+p_{1\mu}\Big\lbrack-2m_S^2(q^2+p_2^2)(m_A^2-p_1^2)-2m_A^2(p_1^2(p_2^2-q^2)
-2q^2p_2^2)\nonumber\\
&&-2 p_1^2 (p_2^4+q^4)
-2 D^{PAS}_\chi p_1^2 (q^2-p_1^2-p_2^2)(q^2-p_2^2)
\Big\rbrack
\Bigg\}
\ea
with
\be
D^{PAS}_{\chi} = i C^{PAS}_\chi \frac{g_S\condc}{g_V\PMNc\vltc}\,.
\ee

\section{Comparison with experiment}
\label{numerics}

The input we use for $\condc$ is the value derived from sum rules
in Ref.~\cite{BPR}, 
which is in agreement with most recent sum rules determinations of this 
condensate and of light quark masses -see \cite{JOPcond02} for instance- 
and the lattice light quark masses world average in \cite{Kaneko02}. 
The value of $F_0$ is from Ref.~\cite{ABT3} and the remaining masses are 
those from the PDG.
\ba
F_0 = (0.087\pm0.006)~\mbox{GeV}\,, && m_V= 0.770~\mbox{GeV}\,,
\nonumber\\
m_A = 1.230~\mbox{GeV} \,,&& m_S = 0.985 ~\mbox{GeV}\,,
\nonumber\\
\frac{\langle \bar u u + \bar d d\rangle^{\overline{MS}}(m_V)}{2}&=& \condc
^{\overline{MS}} (m_V)
= -(0.0091\pm0.0020)~\mbox{GeV}^3\,.
\ea
Putting in the various relations, we immediately obtain
\ba
\label{numresults}
f_V &=& 0.15 ~\lbrack0.20\rbrack ~\cite{ENJLreview,BBR} \,,
\nonumber\\
f_A &=& 0.057 ~\lbrack0.097\pm0.022\rbrack ~\cite{ENJLreview,BBR}\,,
\nonumber\\
L_5 (m_V)&=& 1.95\cdot10^{-3} ~\lbrack(1.0\pm0.2)\cdot10^{-3} \rbrack
~\mbox{\cite{ABT3}} \,,\nonumber\\
L_8  (m_V) &=& 0.5\cdot10^{-3}  ~\lbrack(0.6\pm0.2)\cdot10^{-3}\rbrack 
~\mbox{\cite{ABT3}}\,,\nonumber\\
L_9  (m_V)&=& 6.8\cdot10^{-3}  ~\lbrack(5.93\pm0.43\rbrack\cdot10^{-3}
\rbrack ~\mbox{\cite{BT1}} \,,\nonumber\\
L_{10}  (m_V)&=& -4.4\cdot10^{-3} ~\lbrack(-4.4\pm0.7)\cdot 10^{-3} \rbrack~
\mbox{\cite{BT1,BT2}}\,.
\ea
These numbers\footnote{The value for $L_{10}$ used the
values of $L_9$ from \cite{BT1}, the $2l_5-l_6$ value from \cite{BT2}
and the $p^4$ relation $2l_5-l_6 = 2 L_9+2 L_{10}$.} 
are in reasonable agreement with the experimental values
given in brackets
with the possible exception of $L_5$ which is rather high. We expect to 
have an uncertainty between $30~\%$ and $40~\%$ in our hadronic predictions.
The values in Eq.~(\ref{numresults})
do not depend on the value of the quark condensate.

We cannot determine $\Delta$ at this level. The three-point functions
$PSP$, $VPP$, $SVV$ and $PVA$ can be rewritten
in terms of the above inputs. There is more freedom in those functions
by expanding the underlying $\overline\Pi$ functions to higher order.
These extra terms can usually be determined from the short-distance
constraints up to the problem discussed in Sect.~\ref{SDProblem}.

\section{Difficulties in Going Beyond the One-Resonance Approximation}
\label{trouble}

An obvious question to ask is whether we can easily go beyond the one
resonance per channel approximation used above using the general
resummation based scheme. At first sight one would have said that
this can be done simply by including higher powers in the expansion
of the one-loop two-point functions and/or giving $g_S,g_V$ a momentum
dependence. Since we want to keep the nice analytic behaviour expected
in the large $N_c$ limit with only poles {\em and} have simple
expressions for the one-loop functions and $g_S,g_V$, it turns out to be
very difficult to accomplish. We have tried many variations but essentially
the same type of problems always showed up, related to the fact that
the coefficients of poles of two-point functions obey positivity
constraints. Let us concentrate on the scalar two-point function in the
chiral limit to illustrate the general problem.

In this limit the full two-point function can be written in terms of the
one-loop function as
\be
\Pi_S(q^2) = \frac{\overline\Pi_S(q^2)}{1-g_S\overline\Pi_S(q^2)}\,.
\ee
If we want to give $g_S$ a polynomial dependence on $q^2$ this two-point
function generally becomes far too convergent in the large $q^2$ limit.
The other way to introduce more poles is to expand $\overline\Pi(q^2)$
beyond what we have done before to quartic or higher order.
For the case of two-poles this means we want
\be
1-g_S\overline\Pi_S(q^2) = a (q^2-m_1^2)(q^2-m_2^2)\,.
\ee
However that means we can rewrite
\be
\label{signs}
\Pi_S(q^2) = -\frac{1}{g_S}+\frac{1}{g_S a (m_1^2-m_2^2)}
\left(\frac{1}{q^2-m_1^2}-\frac{1}{q^2-m_2^2}\right)\,.
\ee
{}From Eq.~(\ref{signs}) it is obvious that the residues of the two poles
will have opposite signs, thus preventing this simple approach for
including more resonances. We have illustrated the problem here for the
simplest extensions but it persists as long as both $g_S,g_V$
and the one-loop two-point functions are fairly smooth functions.

\section{A General Problem in Short-Distance Constraints in Higher Green
Functions}
\label{SDProblem}

At this level we have expanded our one-loop two-point functions
to at most second nontrivial order in the momenta and we found that
it was relatively easy to satisfy the short-distance constraints
involving exact zeros. However, if we check the short-distance relations
for the three-point functions that are order parameters given in
Eqs. (\ref{PSPSD}), (\ref{SVVSD}) and (\ref{PVASD}) and compare
with short-distance properties of our model three-point functions of
(\ref{PSPcf}), (\ref{SVVcf}) and (\ref{PVAc}), we find that they
are typically too convergent. In this subsection we will discuss
how this cannot be remedied
in general without spoiling the parts we have already
matched. In fact, we will show how in general this cannot be
done using a single or any finite number of resonances
per channel type of approximations.
An earlier example where single resonance
does not allow to reproduce all short-distance constraints was
found in Ref.~\cite{KN1}.

First look at the function $\Pi^{PSP}$ and see whether by adding terms
in the expansion in $q^2, p_1^2, p_2^2$ to $\overline\Pi^{PSP}(p_1,p_2)^\chi$
beyond those considered in Eq.~(\ref{PSPc}) we can satisfy the short-distance
requirement of Eq.~(\ref{PSPSD}).
It can be easily seen that setting
\ba
\overline\Pi^{PSP}(p_1,p_2)^\chi
&=&
\left.\overline\Pi^{PSP}(p_1,p_2)^\chi\right|_{\mbox{Eq.~(\ref{PSPc})}}
+\overline\Pi_5^{PSP\chi}\left(q^4+p_2^4-2 q^2 p_2^2 -p_1^4\right)\,,
\nonumber\\
\overline\Pi_5^{PSP\chi} &=& 
\frac{\left(\PMNc\right)^3}{16\condc^2\left(1-2g_S g_V\condc\PMNc\right)}
\ea
makes the short-distance constraint Eq.~(\ref{PSPSD}) satisfied.
However, a problem is that now we obtain a very bad short-distance behaviour
for the pion scalar form factor $F_S^\chi(p_1^2)$ which diverges as
$p_1^2$ rather than going to zero.
Inspection of the mechanism behind this shows that this is a general problem
going beyond the single three-point function and model discussed here.

The problem is more generally a problem between the short-distance requirements
on form factors and cross-sections, many of which can be qualitatively
derived from the quark-counting rules or more quantitatively using the methods
of Ref.~\cite{BrodskyLepage}, with the short-distance properties of general
Green functions.

The quark-counting rules typically require a form factor,
here $F_S^\chi(p_1^2)$, to vanish as $1/p_1^2$ for large $p_1^2$.
The presence of the short-distance part proportional to $p_1^2/(q^2 p_2^2)$
in the short distance expansion of $\Pi^{PSP}(p_1,p_2)^\chi$ then
requires a coupling of the hadron in the $P$ channel to the
$S$ current proportional to $p_1^2$ (or via a coupling to a hadron in the
$S$ channel which in turn couples to the $S$ current, this complication
does not invalidate the argument below).
In the general class of models
with hadrons coupling with point-like couplings the negative powers
in Green functions can only be produced by a hadron propagator.
The positive power present in the short-distance expression must thus
be present in the couplings of the hadrons. This in turn implies that this
power is present in the form factor of at least some hadrons. The latter
is forbidden by the quark-counting rule.

It is clear that for at most a single resonance in each channel there is no
solution to this set of constraints. In fact, as will show below,
there is no solution to this problem for any finite number of resonances
in any channel. This shows that even for order parameters the
approach of saturation by resonances might have to be supplemented
by a type of continuum. 
We will illustrate the problem for the PSP three-point function.
The general expression, labeling resonances in the first $P$-channel by
$i$, in the $S$-channel by $j$ and in the last $P$-channel by $k$
is 
\ba
\Pi^{PSP}(p_1,p_2)^\chi &=& 
f_0(q^2,p_1^2,p_2^2)
+\sum_i \frac{f_{1i}(p_1^2,p_2^2)}{(q^2-m_i^2)}
+\sum_j \frac{f_{2j}(q^2,p_2^2)}{(p_1^2-m_j^2)}
+\sum_k \frac{f_{3k}(q^2,p_2^2)}{(p_2^2-m_k^2)}
\nonumber\\&&
+\sum_{ij} \frac{f_{4ij}(p_2^2)}{(q^2-m_i^2)(p_1^2-m_j^2)}
+\sum_{ik} \frac{f_{5ik}(p_1^2)}{(q^2-m_i^2)(p_2^2-m_k^2)}
\nonumber\\&&
+\sum_{jk} \frac{f_{6jk}(q^2)}{(p_1^2-m_j^2)(p_2^2-m_k^2)}
+\sum_{ijk} \frac{f_{ijk}}{(q^2-m_i^2)(p_1^2-m_j^2)(p_2^2-m_k^2)}
\ea
The couplings $f_i$ are polynomials in their respective arguments.
The short-distance constraint now requires
$f_0(q^2,p_1^2,p_2^2) = 0$ and various cancellations between coefficients
of the other functions. The presence of the term $p_1^2/(q^2 p_2^2)$
now requires the presence of at least a nonzero term of order $p_1^2$
in one of the $f_{5ik}(p_1^2)$. However the Green function can then be
used to extract the scalar (transition) form factor between hadron $i$ and $k$
which necessarily increases as $p_1^2$ which is forbidden by the
quark-counting rules for this (transition) scalar form factor.
The terms with $p_2^2/(q^2 p_1^2)$ and $q^2/(p_1^2 p_2^2)$ obviously
leads to similar problems but in other (transition) form factors.

We have discussed the problem here for one specific three-point function
but it is clear that this is a more general problem for three-point
functions. 
For Green function with more than three insertions similar conflicts with the
quark counting rules will probably arise also from hadron-hadron scattering
amplitudes.

\section{Conclusions}

In this paper we have constructed a new approximation to low and intermediate
energy hadronic quantities. Our approach naturally fits in the large
$N_c$ limit and incorporates chiral symmetry constraints by construction.
We have shown that many short-distance constraints can be easily incorporated
but pointed out that our model, but also a more general saturation by
hadrons approach, cannot reconcile all short-distance constraints due
to
a general conflict between short distance constraints on Green functions
and those on form factors and cross-sections
that can be obtained from those Green functions via LSZ reduction.

We have also shown how our approach incorporates the gap equation and
the concept of a constituent quark mass following directly from the
Ward Identities and the resummation assumption.

We have also compared our results with experimental results for hadronic
observables and found reasonable agreement.

\section*{Acknowledgements}
E.G. is indebted to MECD (Spain) for a F.P.U. Fellowship. 
This research is supported in part by the Swedish Research Council,
by MCYT (Spain) Grant No. FPA2000-1558 (E.G and J.P.) and by Junta de 
Andaluc\'{\i}a Grant No. FQM-101 (E.G. and J.P.). 
E.G. and J.P. thank Lund University for hospitality.

\appendix
\renewcommand{\theequation}{\Alph{section}.\arabic{equation}}
\setcounter{equation}{0}
                   
\section{Some Short-Distance Relations Beyond those Mentioned in the Text}

We have calculated or recalculated several short-distance behaviours
of three-point functions.
\be
\label{PSPSD}
\lim_{\lambda\to \infty}\Pi^{PSP}(\lambda p_1,\lambda p_2)^{\chi}
=
\frac{\condc}{2\lambda^2}
\left\{\frac{p_2^2}{q^2 p_1^2}+\frac{q^2}{p_1^2 p_2^2}
-\frac{p_1^2}{q^2 p_2^2}-\frac{2}{p_1^2}\right\}
\ee
\ba
\label{SVVSD}
\lefteqn{
\lim_{\lambda\to\infty}\Pi^{SVV}(\lambda p_1,\lambda p_2)^\chi
= \frac{\condc}{2\lambda^2 q^2 p_1^2 p_2^2}
\Bigg\{ -4 p_2^2\,p_{1\mu} p_{1\nu} - 2 (p_1^2+p_2^2-q^2) \,p_{1\mu} p_{2\nu}
}
&&\nonumber\\&&
-2 (p_1^2 + p_2^2 + q^2) \,p_{2\mu} p_{1\nu} -4 p_1^2\,p_{2\mu} p_{2\nu} 
+ \left( q^4 -(p_1^2-p_2^2)^2 \right)\,g_{\mu\nu}\Bigg\}
\ea
\ba
\label{PVASD}
\lefteqn{
\lim_{\lambda\to\infty}\Pi^{PVA}(\lambda p_1,\lambda p_2)^\chi
 = \frac{i\condc}{2\lambda^2 q^2 p_1^2 p_2^2}
\Bigg\{ 4 p_2^2\,p_{1\mu} p_{1\nu} - 2 (q^2+p_1^2-p_2^2) \,p_{1\mu} p_{2\nu}
}
&&\nonumber\\&&
+2 (q^2 + p_2^2 - p_1^2) \,p_{2\mu} p_{1\nu}  -4 p_1^2\,p_{2\mu} p_{2\nu}
+ \left( p_2^4 -(p_1^2-q^2)^2 \right)\,g_{\mu\nu}\Bigg\}
\ea
Some of these have been mentioned in Refs.~\cite{KN1,Moussallam}.

The following are to our knowledge new:
\ba
\label{PASSD}
\lefteqn{
\lim_{\lambda\to\infty}\left(\Pi^{PAS}(\lambda p_1,\lambda p_2)^\chi
+\Pi^{SAP}(\lambda p_1,\lambda p_2)^\chi\right)
\,=\,
i 4 \pi\alpha_S\frac{N_c^2-1}{N_c^2}\frac{\condc^2}{\lambda^5}}
 \nonumber\\
&&\times\left\lbrace
\frac{2 p_{1\mu}}{p_1^2}
\left(\frac{1}{q^4}+\frac{1}{p_2^4}+\frac{1}{p_2^2 q^2}\right)
+\frac{p_{2\mu}}{p_2^2}
\left(\frac{-1}{q^4}+\frac{1}{p_2^2 q^2}\right)
+\frac{q_{\mu}}{q^2}
\left(\frac{1}{p_2^4}-\frac{1}{p_2^2 q^2}\right)
\right\rbrace
\ea

\ba
\label{PASSDi}
\lim_{\lambda\to\infty}\lim_{m_q\to 0}\frac{\partial}{\partial m_i}
\left(\Pi^{PAS}(\lambda p_1,\lambda p_2)^{ijk}
+\Pi^{SAP}(\lambda p_1,\lambda p_2)^{ijk}\right)
&=&
-2 i  \condc \frac{p_{2\mu}}{\lambda^3 p_2^2 q^2}\,,
\nonumber\\
\lim_{\lambda\to\infty}\lim_{m_q\to 0}\frac{\partial}{\partial m_j}
\left(\Pi^{PAS}(\lambda p_1,\lambda p_2)^{ijk}
+\Pi^{SAP}(\lambda p_1,\lambda p_2)^{ijk}\right)
&=&
2 i  \condc \frac{p_{1\mu}}{\lambda^3 p_1^2}
\left(\frac{1}{q^2}+\frac{1}{p_2^2}\right)\,,
\nonumber\\
\lim_{\lambda\to\infty}\lim_{m_q\to 0}\frac{\partial}{\partial m_k}
\left(\Pi^{PAS}(\lambda p_1,\lambda p_2)^{ijk}
+\Pi^{SAP}(\lambda p_1,\lambda p_2)^{ijk}\right)
&=&
2 i  \condc \frac{q_{\mu}}{\lambda^3 p_2^2 q^2}\,.
\ea


\begin{thebibliography}{99}

\bibitem{largeNc}
G.~'t Hooft,
Nucl.\ Phys.\ B {\bf 72} (1974) 461;\\
A.~V.~Manohar,
hep-ph/9802419,
Les Houches Summer School in Theoretical Physics,
Session 68: Probing the Standard Model of Particle Interactions, Les Houches,
France, 28 Jul - 5 Sep 1997
F. David and R. Gupta eds.

\bibitem{NJL}
Y.~Nambu and G.~Jona-Lasinio,
Phys.\ Rev.\  {\bf 122} (1961) 345.

\bibitem{NJLreviews}
T.~Hatsuda and T.~Kunihiro,
Phys.\ Rept.\  {\bf 247} (1994) 221
[hep-ph/9401310];\\
S.~P.~Klevansky,
Rev.\ Mod.\ Phys.\  {\bf 64} (1992) 649.

\bibitem{ENJLreview}
J.~Bijnens,
Phys.\ Rept.\  {\bf 265} (1996) 369
[hep-ph/9502335].

\bibitem{CQM}
A.~Manohar and H.~Georgi,
Nucl.\ Phys.\ B {\bf 234} (1984) 189;\\
D.~Espriu, E.~de Rafael and J.~Taron,
Nucl.\ Phys.\ B {\bf 345} (1990) 22
[Erratum-ibid.\ B {\bf 355} (1991) 278].

\bibitem{BBR}
J.~Bijnens, C.~Bruno and E.~de Rafael,
Nucl.\ Phys.\ B {\bf 390} (1993) 501
[hep-ph/9206236].

\bibitem{BRZ}
J.~Bijnens, E.~de Rafael and H.~-q.~Zheng,
Z.\ Phys.\ C {\bf 62} (1994) 437
[hep-ph/9306323].


\bibitem{BP1}
J.~Bijnens and J.~Prades,
Z.\ Phys.\ C {\bf 64} (1994) 475
[hep-ph/9403233].


\bibitem{PPR}
S.~Peris, M.~Perrottet and E.~de Rafael,
J.\ High Energy Phys.\  {\bf 05} (1998) 011
[hep-ph/9805442];\\
M.~Knecht, S.~Peris and E.~de Rafael,
Nucl.\ Phys.\ Proc.\ Suppl.\  {\bf 86} (2000) 279
[hep-ph/9910396].

\bibitem{GP}
M.~F.~Golterman and S.~Peris,
Phys.\ Rev.\ D {\bf 61} (2000) 034018
[hep-ph/9908252].

\bibitem{KN1}
M.~Knecht and A.~Nyffeler,
Eur.\ Phys.\ J.\ C {\bf 21} (2001) 659
[hep-ph/0106034].

\bibitem{Moussallam}
B.~Moussallam and J.~Stern,
hep-ph/9404353.

B.~Moussallam,
Phys.\ Rev.\ D {\bf 51} (1995) 4939
[hep-ph/9407402]; 
Nucl.\ Phys.\ B {\bf 504} (1997) 381
[hep-ph/9701400].

\bibitem{BPx}
J.~Bijnens, E.~G\'amiz and J.~Prades,
J.\ High Energy Phys.\ {\bf 10} (2001) 009
[hep-ph/0108240].

J.~Bijnens and J.~Prades,
J.\ High Energy Phys.\  {\bf 06} (2000) 035
[hep-ph/0005189]; 
J.\ High Energy Phys.\  {\bf 01} (2000) 002
[hep-ph/9911392]; 
J.\ High Energy Phys.\  {\bf 01} (1999) 023
[hep-ph/9811472]; 
Nucl.\ Phys.\ B {\bf 490} (1997) 239
[hep-ph/9610360].


J.~Bijnens, E.~Pallante and J.~Prades,
Nucl.\ Phys.\ B {\bf 521} (1998) 305
[hep-ph/9801326].

J.~Bijnens, A.~Fayyazuddin and J.~Prades,
Phys.\ Lett.\ B {\bf 379} (1996) 209
[hep-ph/9512374].



\bibitem{BPPx}
J.~Bijnens, E.~Pallante and J.~Prades,
Nucl.\ Phys.\ B {\bf 626} (2002) 410
[hep-ph/0112255]; 
Nucl.\ Phys.\ B {\bf 474} (1996) 379
[hep-ph/9511388]; 
Phys.\ Rev.\ Lett.\  {\bf 75} (1995) 1447
[Erratum-ibid.\  {\bf 75} (1995) 3781]
[hep-ph/9505251].

J.~Bijnens and J.~Prades,
Nucl.\ Phys.\ B {\bf 444} (1995) 523
[hep-ph/9502363]; 
Phys.\ Lett.\ B {\bf 342} (1995) 331
[hep-ph/9409255].


\bibitem{Weinberg}
S.~Weinberg,
Phys.\ Rev.\ Lett.\  {\bf 18} (1967) 507.

\bibitem{RRY}
M.~A.~Shifman, A.~I.~Vainshtein and V.~I.~Zakharov,
Nucl.\ Phys.\ B {\bf 147} (1979) 385;\\
L.~J.~Reinders, H.~Rubinstein and S.~Yazaki,
Phys.\ Rept.\  {\bf 127} (1985) 1.

\bibitem{BrodskyLepage}
G.~P.~Lepage and S.~J.~Brodsky,
Phys.\ Lett.\ B {\bf 87} (1979) 359.

G.~P.~Lepage and S.~J.~Brodsky,
Phys.\ Rev.\ D {\bf 22} (1980) 2157.

\bibitem{JOP01}
M.~Jamin, J.~A.~Oller and A.~Pich,
Nucl.\ Phys.\ B {\bf 622} (2002) 279
[hep-ph/0110193].
A.~Pich,
[hep-ph/0205030].

\bibitem{BPR}
J.~Bijnens, J.~Prades and E.~de Rafael,
Phys.\ Lett.\ B {\bf 348} (1995) 226
[hep-ph/9411285];\\
K.~Maltman and J.~Kambor,
Phys.\ Lett.\ B {\bf 517} (2001) 332
[hep-ph/0107060].

\bibitem{JOPcond02}
M.~Jamin, J.~A.~Oller and A.~Pich,
Eur.\ Phys.\ J.\ C {\bf 24} (2002) 237
[hep-ph/0110194].

\bibitem{Kaneko02}
T.~Kaneko,
Nucl.\ Phys.\ Proc.\ Suppl.\  {\bf 106} (2002) 133
[hep-lat/0111005].

\bibitem{ABT3}
G.~Amor\'os, J.~Bijnens and P.~Talavera,
Nucl.\ Phys.\ B {\bf 602} (2001) 87
[hep-ph/0101127].

\bibitem{BT1}
J.~Bijnens and P.~Talavera,
J.\ High Energy Phys.\  {\bf 03} (2002) 046
[hep-ph/0203049].

\bibitem{BT2}
J.~Bijnens and P.~Talavera,
Nucl.\ Phys.\ B {\bf 489} (1997) 387
[hep-ph/9610269].
\end{thebibliography}
\end{document}